# Thermoelectric power factor of nanocomposite materials from two-dimensional quantum transport simulations

Samuel Foster[1*], Mischa Thesberg[2], and Neophytos Neophytou[1]

[1]School of Engineering, University of Warwick, Coventry, CV4 7AL, UK

[2]Institute for Microelectronics, TU Wien, Vienna, Austria, A-1040

[*] S.Foster@warwick.ac.uk

## Abstract

Nanocomposites are promising candidates for the next generation of thermoelectric materials since they exhibit extremely low thermal conductivities as a result of phonon scattering on the boundaries of the various material phases. The nanoinclusions, however, should not degrade the thermoelectric power factor, and ideally should increase it, so that benefits to the *ZT* figure of merit can be achieved. In this work we employ the Non-Equilibrium Green's Function (NEGF) quantum transport method to calculate the electronic and thermoelectric coefficients of materials embedded with nanoinclusions. For computational effectiveness we consider two-dimensional nanoribbon geometries, however, the method includes the details of geometry, electron-phonon interactions, quantisation, tunneling, and the ballistic to diffusive nature of transport, all combined in a unified approach. This makes it a convenient and accurate way to understand electronic and thermoelectric transport in nanomaterials, beyond semiclassical approximations, and beyond approximations that deal with the complexities of the geometry. We show that the presence of nanoinclusions within a matrix material offers opportunities for only weak energy filtering, significantly lower in comparison to superlattices, and thus only moderate power factor improvements. However, we describe how such nanocomposites can be optimised to limit degradation in the thermoelectric power factor and elaborate on the conditions that achieve the aforementioned mild improvements. Importantly, we show that under certain conditions, the power factor is independent of the density of nanoinclusions, meaning that materials with large nanoinclusion densities which provide very low thermal conductivities, can also retain large power factors and result in large *ZT* figures of merit.

# I. Introduction

Thermoelectric materials convert heat from temperature gradients into electrical voltages and vice versa. The performance of such materials is quantified by the dimensionless figure of merit: $ZT = \sigma S^2 T/(\kappa_e + \kappa_\ell)$ where $\sigma$ is the electrical conductivity, $S$ is the Seebeck coefficient, $T$ is the temperature, $\kappa_e$ is the electronic thermal conductivity, $\kappa_\ell$ is the lattice thermal conductivity, and $\sigma S^2$ is known as the power factor ($PF$). Traditionally, $ZT$ has been approximately 1 in the limited range of materials used in commercial applications, which are mostly semiconductor doped alloys of Antimony and BiTe at room temperature[1], and PbTe or SiGe at higher temperatures[2].

More recently numerous other bulk materials have been studied or characterized such as transition metal dichalcogenides (TMDC),[3-6] skutterudites,[7-9] phonon-glass-electron crystal structures,[10] and half-Heuslers.[11-13] A large number of these materials demonstrate $ZT$ above 1, primarily by the reduction of the thermal conductivity, $\kappa$.[14]

Many methods also exist for reducing $\kappa$ beyond bulk values. These include superlattices,[15] alloying,[16] heavy doping,[17] nanoporous materials,[18-20] and nanograining.[21,22] One of the most widespread methods for the reduction of the thermal conductivity has been the use of nanoinclusions.[23-28] These cause scattering of short wavelength phonons and can produce significant reductions in $\kappa$. This is because in common thermoelectric materials, such as PbTe, a large portion of the phonons have mean-free-paths for scattering on the order of nanometers.[29] This technique is therefore widely used to enhance thermoelectric performance in a broad range of materials, including BiTe,[30,31] PbTe,[23,32,33] SiGe,[16,34,35] ZnSb,[25,36] FeSi,[37] MnSi,[38] SnTe,[39] PbS,[40] CuSe,[41] YbCoSb,[42] and ZrNiSn.[43] Indeed, by embedding nanoinclusions within PbTe in a hierarchical manner, record high $ZT = 2.2$ values were achieved due to drastic reductions in $\kappa$, but also due to retaining high power factors.[23] Ref. 44, in particular, denotes the importance of matrix/inclusion band alignment to retain the original conductivity of the material and avoid degradation in the power factor.

While the impact of nanoinclusions on the thermal conductivity is well documented,[18,45] previous works are not as clear on their impact on the power factor, with results varying significantly, from only small influence,[30,31,34,46] to large potential improvements.[25,36,42,47] Thus, it is imperative that a high level of understanding on the influence of nanoinclusions on the power factor, both qualitative and quantitative is also established, if $ZT$ is to be maximized. However, the complexity of the electronic transport, combining semiclassical effects, quantum effects, ballistic and diffusive regimes, as well as the geometry details, makes accurate modelling a difficult task. Several works in the literature use semi-classical models, simplified geometries, and various approximations to provide understanding of transport in such systems.[28,46-48]

In this work we show how the Non-Equilibrium Green's Function (NEGF) simulation method can be employed to calculate the electron transport properties in 2D nanostructures in a fully quantum mechanical way that includes the details of geometry, electron-phonon interactions, quantisation, tunneling, and the ballistic to diffusive nature of transport, all combined in a unified simulation approach. Such simulations are very demanding, thus, for computational effectiveness we consider 30 nm × 60 nm 2D nanoribbon channels embedded with nanoinclusions in a regular hexagonal configuration. These short channels are, however, large enough to capture all essential transport physics as we will explain. We present a detailed study of the influence of nanoinclusions on the $PF$ of nanocomposite materials. We show that, unfortunately, the presence of potential barriers originating from nanoinclusions within a matrix material offers opportunities for only moderate power factor improvements, resulting from their inability to act as effective energy filters, a behavior very different than that of superlattice structures. We describe, however, how such nanocomposites can be optimised to limit $PF$ degradation and even achieve mild improvements. We show that the key design elements for this $PF$ resilience is to begin with a degenerately doped matrix material in which the Fermi level is placed 1-$2k_BT$ into the bands, and then insert nanoinclusions of barrier heights between the Fermi level and conduction band edge. This introduces a small filtering effect which improves the Seebeck coefficient and is more effective when the nanoinclusions are large enough to prevent quantum tunneling. Importantly, we also show that under these conditions, the power factor is independent of the density of nanoinclusions, even slightly benefiting at higher densities (where strong reduction in $\kappa$ is also anticipated). This provides opportunities for dense nanoparticle materials with low $\kappa$ and still acceptable $PF$s, thus high $ZT$ figures of merit.

Thus, the goal of this work is to illuminate aspects of the thermoelectric power factor in nanostructures for which several contradicting reports are encountered in the literature. The paper is organised as follows: In Section II we describe our NEGF approach including our calibration procedure and indicate the geometries we study. In Section III we present our results. In Section IV we discuss the results, and in Section V we conclude.

## II. Approach

To compute the electronic transport, we have developed a 2D quantum transport simulator based on the Non-Equilibrium Green's Function (NEGF) formalism including electron-acoustic phonon scattering in the self-consistent Born approximation.[49,50] This approach can capture all relevant quantum effects such as quantization, energy mixing, interferences, and tunneling, as well as all geometrical complexities, which can be important in transport through disordered materials.

The system is treated as a 2D channel within the effective mass approximation, where we use a uniform $m^* = m_0$ in the entire channel, where $m_0$ is the rest mass of the electron. The nanoinclusions are modeled as potential barriers of cylindrical shape within the matrix material as shown in the schematic of Fig. 1c. We consider regular hexagonal placement of the nanoinclusions, but in the discussion section we elaborate on the possible effects of their random placement based on our findings. The NEGF theory is described adequately in various places in the literature[49-51] so we do not include it here. Most work on NEGF in the literature is applied to 1D systems due to computational limitations, however in this work we expand the formalism to 2D systems of widths $W = 30$ nm and lengths $L = 60$ nm (see Fig. 1c). The Recursive Green's Function (RGF) formalism is used to calculate the relevant elements of the Green's function, and the Sancho-Rubio algorithm to compute the self-energies of the contacts.[52]

The effect of electron scattering with acoustic phonons in NEGF is modeled by including a self-energy on the diagonal elements of the Hamiltonian. This approximation has been shown to be quantitatively valid for many systems,[53] such as electrons in silicon,[54] transport in carbon nanotubes,[49] and many more, and captures the essential transport features. The convergence criteria for the ensuing self-consistent calculation is chosen to be current conservation, i.e. we consider convergence is achieved when the current is conserved along the length of the channel to within 1%. The strength of the electron-phonon coupling is given by $D_0$, which we consider uniform across the entire channel. This parameter, which has units of $eV^2$, represents the weighting of the Green's Function contributions to the scattering self-energy. Its relation to the deformation potential can be found in Refs. 49 and 55.

The power factor, $GS^2$, is obtained using the expression:

$$I = G\Delta V + SG\Delta T. \qquad (1)$$

For each value of the power factor, the simulation is run twice, initially with a small potential difference and no temperature difference ($\Delta T$=0), which yields the conductance ($G = I_{(\Delta T=0)}/\Delta V$), then again with a small temperature difference and no potential difference ($\Delta V$=0), which yields the Seebeck coefficient ($S = I_{(\Delta V=0)}/G\Delta T$). This method is validated in Ref. 56. The sharp features of the system required a large number (~100) of convergence steps. Figure 1c shows a typical band diagram of the nanocomposite under consideration. The Fermi level is denoted by the dashed-red line. Current flows through the nanoinclusion barriers and over them.

**Channel calibration**: Previous theoretical and experimental works[21,22,55,57,58] have shown degenerately doped materials, once nanostructured to improve filtering, could provide significant power factor increases. Placing the Fermi level well into the bands improves conductivity, which compensates for the reduction that is caused by nanostructuring. Thus, in this work as well, as a starting point, we place the Fermi level high into the bands at $2k_BT$ above the conduction band edge. We assume room temperature $T = 300$ K throughout the paper. The value of $D_0$ is then chosen such that the conductance of an $L = 15$ nm long pristine channel is found to be 50% of the ballistic value. This effectively amounts to fixing a mean-free-path of 15 nm for the system; a value that is comparable to common semiconductors such as silicon.[59-61] The appropriate $D_0$ was found to be $D_0 = 0.0026$ $eV^2$ as shown in Fig. 1a. Thus, with such a mean-free-path, the $L = 60$ nm channel length we consider is large enough to result in diffusive transport in the material we simulate, although in the discussion section we also elaborate on the features of ballistic transport. The conduction band is set at $E_C = 0.00$ eV and the Fermi level, unless otherwise stated, is placed at $E_F = 0.05$ eV. It should be noted that the chosen value of $D_0$ only produces a mean-free-path (as defined here) of exactly 15 nm when $E_F = 0.05$ eV as this is the Fermi level used during the calibration. As the Fermi level moves, the average energy of the electrons changes and consequently so does the mean-free-path, deviating somewhat linearly as the $E_F$ changes. We can then extract the power factor as shown in Fig. 1b versus the reduced Fermi level $\eta_F$, i.e. the position of the Fermi level with respect to the band edge, $\eta_F = (E_F\text{-}E_C)/k_B T$. As expected, the maximum power factor is observed when the Fermi level is in the vicinity of the band edge.[62]

With regards to the transport properties, in Fig. 1d we show the transmission function of the nanocomposite channel under four different conditions: i) coherent (ballistic) transport for a pristine channel (blue 'staircase' line), ii) coherent transport for a channel with nanoinclusions (light-blue line), iii) incoherent transport for a pristine channel (red line), and iv) incoherent transport for a channel with nanoinclusions (light-red line). The barrier height of the nanoinclusions is set to $V_B = 0.01$ eV and the Fermi level at $E_F = 0.05$ eV. The ballistic transmission of the pristine channel shows the usual staircase shape, with an increment every time a new subband is reached in energy. A large drop is observed when the nanoinclusions are added in the geometry, where resonance features are also evident. Those features are removed when phonon scattering is included, and the transmission is reduced even more when nanoinclusions are added in addition to phonon scattering.

An interesting feature from these results is the fact that the transmission suffers significantly once the nanoinclusions are added, even at energies much higher than the barrier height, and we elaborate on this more in the Discussion Section IV. This is in contrast to a common approximation that energies above the barrier are not severely affected and are considered to be restored to their pristine material value. The transmission in this case is dominated by the regions of high resistance, which are the nanoinclusions. In the nanoinclusion regions, the bands that contribute to transmission begin just above $V_B$, i.e. it is as if the ballistic transmission is shifted downwards by the number of bands it has at $V_B$. Since in 2D there are numerous numbers of subbands at lower energies, the reduction in the transmission is strong, and it is not recovered even at energies much higher that $V_B$.

# III. Results

Once the calibration is completed we proceed to consider geometries which include circular nanoinclusions (NIs) of different barrier heights, $V_B$, different NI densities, and different NI diameters. The channel width was kept at $W = 30$ nm, and the length at $L = 60$ nm in all cases.

**<u>Influence of barrier height $V_B$ and Fermi level position $E_F$</u>:** The first investigation we perform is on the influences of: i) the nanoinclusion barrier height $V_B$, and ii) the Fermi level, $E_F$, on the thermoelectric coefficients, conductance $G$, Seebeck coefficient $S$, and power factor $GS^2$. Transport in an 8×4 hexagonal array of nanoinclusions of diameter $d = 3$ nm (as indicated in the inset of Fig. 2c) is simulated at five different Fermi levels, $E_F = -0.025$ eV (purple lines), $E_F = 0$ eV (green lines), $E_F = 0.025$ eV (black lines), $E_F = 0.05$ eV (red lines), and $E_F = 0.075$ eV (blue lines). For each Fermi level, we vary the nanoinclusion barrier height from $V_B = 0$ eV to $V_B = 0.1$ eV in steps of 0.01 eV. These are similar band offset values that one encounters in promising thermoelectric materials, for example, PbSe/CdSe with a valence band offset of 0.06 eV, PbSe/ZnSe with a valence band offset of 0.13 eV, and PbS/CdS with a valence band offset of again 0.13 eV.[44] The comprehensive results are shown in Fig. 2a, 2b, and 2c for the conductance $G$, the Seebeck coefficient $S$, and the power factor $GS^2$, respectively. As can be observed in Fig. 2a, the conductance $G$ shows the expected decrease at all Fermi levels as $V_B$ is increased, due to the potential barriers blocking the electron flow. For higher barriers $G$ saturates, with the saturation being observed more evidently ~$2k_BT$ above the Fermi level, i.e. the saturation tends to shift to the right with increasing $E_F$. Increasing the Fermi level increases the conductance as well, since higher velocity states are increasingly occupied. Naturally, as the Fermi level increases, the Seebeck coefficient in Fig. 2b drops almost linearly (comparing the different lines in Fig. 2b) following the usual reverse trend compared to $G$. The Seebeck coefficient is proportional to the average energy of the current flow with respect to the Fermi level, $S \propto \langle E - E_F \rangle$ which is reduced as the Fermi level is raised until degenerate conditions are reached. At each individual constant Fermi level line, the Seebeck coefficient only slightly increases with $V_B$, a sign of weak energy filtering, before it saturates as also observed in the case of $G$.

The corresponding power factors are shown in Fig. 2c. Comparing the lines that correspond to the various Fermi levels, a large variation in the power factor is observed in the left of Fig. 2c, for small nanoinclusion barrier heights. As $V_B$ increases, the power factors follow a declining trend and finally all lines saturate at a lower value compared to the pristine material power factors. One important observation that can be detected from Fig. 2c is that the highest power factor is observed for the channel where the Fermi level is placed around the conduction band edge, or somewhat higher (green and black lines, $E_F$ = 0 eV, 0.025 eV), but more importantly when the band edges of the matrix and the nanoinclusions are aligned (i.e. $V_B$ = 0 eV). This clearly shows that in principle the introduction of energy filtering potential barriers by the use of nanoinclusions *cannot* increase the power factor. This is of course if one considers a material with an optimized Fermi level position at $E_F \sim E_C$ to begin with, which is rarely the case in practice. If one considers, however, that the position of the Fermi level is in general not at the optimal point, then there is a possibility of moderate power factor improvements of the order of ~10% (red, blue lines). The power factor lines in Fig. 2c for $E_F > E_C$ indicate that as the barrier heights $V_B$ of the nanoinclusions increase, a maximum is reached when $V_B$ is approximately halfway between $E_F$ and $E_C$, producing a 5-10% increase in the power factors. Raising $V_B$ even further takes away this increase and forces the power factor to saturate at a lower level (to around 50% of the initial PF). This requirement for small band offsets to retain high conductivity has previously been identified in Refs. 14, 44, 63, 64 but it's effect on the power factor had not yet been quantified.

**Influence of the nanoinclusion density:** The next investigation we perform is to illustrate the influence of the NI density on the thermoelectric coefficients. Fig. 3 shows the thermoelectric coefficients $G$, $S$ and $PF$, again versus nanoinclusion barrier height $V_B$ for four different geometries of increasing density as shown in the insets of Fig. 3c. These four simulated geometries are: a 2×4 array (green lines), a 4×4 array (black lines), a 6×4 array (blue lines), and an 8×4 array (red lines). The Fermi level is again placed at $E_F$ = 0.05 eV (dashed-red line in Fig. 3c). Fig. 3a shows that, as before, the conductance $G$ falls as $V_B$ increases, and, as expected, $G$ also falls as the number of nanoinclusions in the channel is increased. Likewise, as the number of nanoinclusions increases, the effect of energy filtering is increased and an improvement in $S$ is observed. The increase is of the order of 10% for the 2×4 channel, and is increased to approximately 25% for the 8×4 channel as seen in Fig. 3b. As $V_B$ increases, we initially see a linear rise in $S$. At barrier heights $V_B$, somewhere between the conduction band edge and the Fermi level, $S$ peaks. For larger $V_B$ it decreases slowly before saturating for barrier heights much above the Fermi level. It is interesting to observe that in this region, both $G$ and $S$ are simultaneously decreasing, a counterintuitive effect – we provide an explanation for this later. Fig. 3c shows the result of these features on the power factor. From zero barrier heights up until $V_B \sim k_BT$, a small increase in the power factors is observed, with a maximum of the order of 10% for the 8×4 channel (red line). As $V_B$ increases even further, the power factor then falls to values below the pristine channel value for all channels. Although for small barrier heights of $V_B < k_BT$ the density has little effect on the power factor, the fact that the $PF$ increases, and is even independent of NI density, is quite important. It indicates that the density of nanostructured materials with nanoinclusions can be optimised for maximal reduction in the thermal conductivity, fine-tuning the distances between the nanoinclusions in order to be of the length scale of the phonon mean-free-path without causing any adverse effects on the power factor. At higher $V_B$, on the other hand, the detrimental effect of density is more important, with the decrease from pristine material power factor ranging from 17% for the 2×4 array to 40% for the 8×4 array as the barrier height is increased to $V_B = 0.1$ eV.

The results in Fig. 2 and Fig. 3, indicate that although the possibility of using nanocomposites with nanoinclusions embedded within a matrix material to improve the power factor is limited, importantly, neither will the careful use of such nanoinclusions limit the power factor significantly. The main reason for using such nanostructures is to provide $ZT$ improvements by reducing the thermal conductivity of the material, and the results of Fig. 2 and Fig. 3 show that such structures can provide resilience to the power factor, as well as showing the possibility of slight benefits. Note here that in our simulations we only consider acoustic phonon scattering. The presence of impurity scattering as well as variation in the thermal conductivities of the different regions can also improve the Seebeck coefficient even further as explained in Refs. 21, 58, 65, 66, which might allow for higher power factors compared to what we compute here.

A non-intuitive point to elucidate here, is the simultaneous drop in both $G$ and $S$ as the barrier height $V_B$ of the nanoinclusions is increased. What is non-intuitive is that in general these two quantities follow a reverse trend, i.e. as $G$ is decreased at the presence of large $V_B$, we would have expected $S$ to increase. The fact that both quantities drop causes a large degradation to the power factor, and limits the filtering capabilities of such nanocomposites. To understand this simultaneous decrease we must consider what happens to the average energy of the current flow as $V_B$ increases, since this determines the Seebeck coefficient $S \propto \langle E - E_F \rangle$.[56] The x-axis of Fig. 4 shows the distribution of the energy of the current flow, $E \times I(E)$, with the average marked with a star, for six different barrier heights, $V_B = 0$ eV (black line), $V_B = 0.02$ eV (red line), $V_B = 0.04$ eV (blue line), $V_B = 0.06$ eV (green line), $V_B = 0.08$ eV (purple line), $V_B = 0.1$ eV (brown line). The inset of Fig. 4 zooms around the average energy of the current flow. As $V_B$ is initially raised, some of the lower energy electrons are cut off while higher energy electrons are less affected, raising the average energy of the current, and thus, raising the Seebeck coefficient (see from black, to red, to blue lines in the inset of Fig. 4). This behavior continues as long as $V_B$ is below the $E_F$, i.e. $V_B < E_F$. It is important to note that electrons with energies less than the barrier height can still contribute to the current by flowing around the nanoinclusion barriers (which is a different scenario compared to superlattice structures which are commonly employed for thermoelectric energy filtering). Thus, as $V_B$ continues to increase, lower energy electrons continue to flow around the barriers and so their contribution is hardly affected because the change in the barrier affects only much larger energies. Higher energy electrons, however, are then begun to be cut off and the average energy begins to fall again (see from blue, to green, to purple, to brown line in the inset of Fig. 4), as does the Seebeck coefficient. Eventually, $V_B$ is high enough that it affects all energies similarly, and the majority of the current flows around the nanoinclusions. Therefore, additional increases in $V_B$ have little effect, and the Seebeck coefficient saturates.

**Influence of quantum tunneling:** An important effect that needs to be considered in evaluating the influence of nanoinclusions on the power factor of nanocomposites is quantum mechanical tunneling. In prior works related to the effect of tunneling in superlattices, we have shown that tunneling is detrimental to the Seebeck coefficient as it makes potential barriers transparent and takes away any benefits that energy filtering could provide to the power factor.[65,66] We have shown that in the case of superlattices tunneling becomes important when the thicknesses of the barriers drop below 1-2 nm (for channels with effective mass $m^* = m_0$). Here, we compare the case of nanoinclusions of small diameters $d \sim 1$ nm which would be strongly influenced by tunneling, versus the case of structures with larger diameters $d \sim 3$ nm, which we expect not to be influenced by tunneling to such a degree. Figure 5 shows the effect of nanoinclusion diameter on the thermoelectric coefficients $G$, $S$ and $PF$ for the $d = 1.5$ nm nanoinclusions (red lines, S1) and $d = 3$ nm nanoinclusions (black lines, S2) for the geometry with 8×4 nanoinclusion arrays (first two insets of Fig. 5c). As before, $G$ falls with increasing $V_B$ in both diameter cases, but the fall is more marked for nanoinclusions of larger diameters, which hinder transport more (Fig. 5a). The smaller diameter nanoinclusions not only occupy less space that obscures transport, but quantum tunnelling causes them to become semi-transparent and allow some current to flow through them. Likewise, due to their transparent nature they do not cause large changes in the Seebeck coefficient as shown by the red line in Fig. 5b (only a ~5% increase is observed at high $V_B$), thus, only a weak energy filtering effect is observed. Consequently, the power factor results in Fig. 5c for the $d = 1.5$ nm nanoinclusions do not show any beneficial effect on the power factor for any of the barrier heights. The beneficial effects of energy filtering are only seen for the larger diameter of $d = 3$ nm, although, as explained earlier, this only appears to occur up to a barrier height approximately halfway between $E_C$ and $E_F$. Beyond this $V_B$, the degradation in $G$ outweighs the gains in $S$, and the power factor falls even further below than that of the NIs with diameter $d = 1.5$ nm. These results demonstrate that, as with superlattices,[65,66] quantum tunnelling has a detrimental impact on the energy filtering effect and, thus, on any potential Seebeck coefficient improvements. To prevent this, diameters of $d > 3$ nm should be used to obtain power factor enhancements (the diameters of course need to be adjusted according to the effective mass of the carriers in the specific material under consideration).

In order to further understand the influence of tunneling versus density of nanoinclusions, in Fig. 5 we also plot the situation where we keep the areal density of nanoinclusions the same as that of the $d = 3$ nm 8×4 array channel (S2), using a lot more nanoinclusions of diameter $d = 1.5$ nm as shown in the third inset of Fig. 5c. Now we have a 15×7 array channel (S3) where the total area of included material is approximately the same across the two structures. The thermoelectric coefficients for this case are shown in Fig. 5 by the blue lines. Quite interestingly, this channel behaves very close to the $d = 3$ nm 8×4 array channel, indicating that at first order one can consider that the overall areal density of nanoinclusions has a stronger impact in determining the thermoelectric performance, compared to the actual size and their distribution. Although the $d = 1.5$nm nanoinclusions will still be semi-transparent, in this case they are many, and are placed in close proximity, in distances smaller than the carriers' relaxation length. This introduces quantum reflections and interferences, which introduce a larger resistance (lower $G$) and increase the energy filtering effect (higher $S$). Figures 6a and 6b show the transmission probability versus energy for the three channels at low $V_B = 0.02$ eV and high $V_B = 0.07$ eV. Indeed, the transmissions of this channel (blue lines) follow closely those of the channel with large $d = 3$nm NIs (black lines). A higher Seebeck coefficient is observed for this channel (blue line) at higher $V_B$ in Fig. 5b because the transmission shows sharper variations around the Fermi level (Fig. 6b). If now one looks carefully back in the $PF$ results of Fig. 5c (blue line), it can be seen that such a channel is the worst of both previously examined channel cases, with no noticeable power factor improvement for low $V_B$ (in contrast to what is shown by the black line), and large $PF$ degradation at high $V_B$ (even stronger than what is shown by the black line). Thus, an important recommendation at this point, is that nanoinclusions with low barrier heights and larger diameters are preferable for power factor resilience.

## IV. Discussion

**<u>Features of the electron flow</u>**: To better understand the electronic transport and transmission (as previously shown in Fig. 1d) through the structures we consider, we show in Fig. 7a a colour plot of the component of the current flow along the length of the structure. Results are taken from the $d = 3$ nm 8×4 channel with $V_B = 0.02$ eV and $E_F = 0.05$ eV. The blue regions indicate the nanoinclusions (where through them the current is low), whereas the yellow regions indicate the matrix material (where the current is high). Note that this spatially varying current is still conserved in the transport direction at all energies independently, i.e. if we integrate along the width direction at every point along the length we get the same value. In Fig. 7b we show a cross-section of the $L$-directed current through two of the nanoinclusions (shown by the dashed-black line in Fig. 7a) at four different energies: $E = 0.01$ eV (below the $V_B$, green line), $E = 0.02$ eV (at the $V_B$, black line), $E = 0.05$ eV (at the $E_F$, blue line), $E = 0.075$ eV (above the $E_F$, purple line). From Fig. 7a it can be seen that the current is reduced where the nanoinclusions are situated (blue areas), but the area affected by the nanoinclusions is not quite the same as the nanoinclusion itself. Due to quantum tunneling, the sides of the nanoinclusion are semi-transparent, narrowing the affected area, while in the direction of current flow, the affected area is elongated due to reflections off the nanoinclusion face. This can also be seen in Fig. 7b where there is a dip in the current at the position of the nanoinclusion and beyond. Crucially, this occurs at all energies where current is still flowing, showing that electrons with energies much higher than $V_B$ are still significantly affected as they pass over the barrier. More detail on this is given in Fig. 7c where we plot the current as it varies in energy at two different points in the channel: i) at the centre of one of the nanoinclusions (blue line, position shown by the dotted-blue line in Fig. 7b), and ii) in the pristine matrix material (black line, position shown by the dotted-black line in Fig. 7b). The barrier height is shown by the dashed-black line and Fermi level by the dashed-red line. It might have been expected that flow below $V_B$ would be cut off and flow above it unaffected. What we see from Fig. 7c however, is that current still flows through the nanoinclusion at lower energies by quantum tunnelling, and at higher energies (even as high as ~2 $k_BT$ above $V_B$) the current has not yet reached the pristine matrix material level. Due to this far-reaching effect of the nanoinclusion, it also appears that there is no clear relation between the optimal $V_B$ and the position of $E_F$ in the results above, other than the optimal $V_B$ for maximizing the power factor appears to be approximately half-way between the band edge and the Fermi level. We next discuss this effect with comparison to superlattices.

**Nanoinclusions vs Superlattices (SLs) - transport features:** Other than the reduction of thermal conductivity, the incorporation of nanoinclusions would have been expected to provide an energy filtering effect and consequently improve the power factor as is the case in transport through cross-plane superlattices (SLs) composed of potential barriers and wells. In SLs, the electrons in the wells have to overpass the heights of the barriers. The higher the barrier, the stronger the reduction in the conductance, which overall is exponential in nature, whereas the Seebeck coefficient increases linearly with the barrier height. It is interesting to compare how the presence of nanoinclusions and superlattice potential wells each influence electronic and thermoelectric transport. In Fig. 8 we plot the transmission of a $L = 60$ nm channel under ballistic coherent conditions for three cases as shown in the insets: i) pristine channel (red line), ii) channel with an 8×4 hexagonal array of NIs (blue line) with barrier height $V_B = 0.1$ eV and diameter $d = 3$ nm, and iii) a SL structure of 8 barriers of height $V_B = 0.1$ eV and width $W = 3$ nm (black line). What is important to note is the differing effects the two structures have on the low energy electrons below the barrier height. The SL structure effectively cuts of the current flow below $V_B = 0.1$ eV, providing an energy filtering mechanism that increases the Seebeck coefficient. The behavior in the presence of NIs is different, because the charge carriers flow not only above the NI barriers, but in between them as well. This means the NIs still allow a finite transmission of carriers across low energies, and thus, do not provide the energy filtering effect and large Seebeck coefficients that can be achieved in superlattices.[58,65,66]

At higher energies, however, the current does not return to the ballistic value in either the nanoinclusion or the superlattice case, in contrast to what is normally assumed. This might explain why improvements in the power factor from superlattices have yet to be realized, as the conductivity falls further than expected with increasing barrier height. Note that this is an effect that originates from the large mismatch between the number of bands in the matrix material and the barrier, and due to the large degree of quantum interferences. Thus, we expect this to be stronger in 2D, compared to 1D where only one (or fewer) subbands exist in all regions of the structure, for example. We also note that simplified models that consider simple step-function-like transmissions (or even simple 1D transmissions) would provide larger conductance and overestimate the performance. However, in the case where incoherent scattering is stronger, this effect would be reduced.

As a comparison between the PF improvements in the two geometries, however, in a superlattice, the power factor can be optimized by placing the Fermi level high into the conduction band (achieving good conductance). The introduction of the barriers increases the Seebeck coefficient by using barriers $\sim k_B T$ above $E_F$, and finally power factor improvements of the order of $\sim$10-20% can be achieved.[56] In the case of channels with nanoinclusions, on the other hand, as shown in Fig. 2c, due to the limited increase in $S$ achieved with nanoinclusions, a somewhat lower power factor enhancement is produced. For non-degenerate conditions ($E_F$ = -0.025 eV and $E_F$ = 0 eV) the conductance drops faster than the Seebeck coefficient rises, and the nanoinclusions have no beneficial effect on the power factor. For degenerate conditions ($E_F$ = 0.025 eV, $E_F$ = 0.05 eV and $E_F$ = 0.075 eV) there is an initial benefit, but in principle, power factor enhancements beyond the pristine structure (with $V_B$ = 0 eV and $E_F$ = 0 eV), are not achieved.

**Random variations in nanoinclusion parameters:** In this work we exclusively considered structures in which the nanoinclusion geometry, diameter, and density were set in a very specific way, i.e. regular hexagonal arrays of fixed diameter. In reality the nanostructuring in nanocomposite materials takes random forms. The specific location of nanoinclusions, their size, the barrier height and even their density cannot be controlled precisely. Even the position of the Fermi level $E_F$, which is set by the doping cannot be controlled precisely. In superlattices, for example, we have shown in a previous work that variations in the lengths of the various regions do not affect the power factor significantly, however what is detrimental are variations in the barrier heights (that degrade the conductivity) and extremely thin, easy to tunnel barriers (which degrade the Seebeck coefficient).[65,66] Although in this work we do not perform a full investigation of the influence of statistical variations of the different structure parameters, from the results in Fig. 2, Fig. 3, and Fig. 5, we can extract some expectations on the effect of variations. If we concentrate at the low $V_B$ regions, where the power factor does not suffer, the results in Fig. 3 indicate that variability in the nanoinclusion density does not affect the power factor, which indicates that variability in the geometry and positions of the nanoinclusions will also not affect the power factor. Interestingly, the results seem tolerant to significant changes in $V_B$, which indicates that moderate barrier height variations will also not affect the power factor either, in contrast to the superlattice case. In superlattices variability in the barrier heights is crucial because carriers need to go through each individual barrier, and the height degrades the conductivity exponentially, whereas in the NI case carriers can actually flow around the nanoinclusions and avoid large barriers. From Fig. 5 we can also observe that quantum tunneling is not as important either, as the energy filtering capabilities of nanoinclusions are limited anyway (in the case of superlattices energy filtering is strong, and tunneling by making the barriers transparent takes it away). From Fig. 2 we can see that the only significant variation that can affect the power factor of the nanocomposite at the low $V_B$ region is the position of the Fermi level $E_F$, which, however, is the case in all materials, nanostructured or not. Another important variability case that is beneficial to the power factor is the variation in the lattice thermal conductivity between the different materials that form the nanocomposite. In superlattices, for example, a lower lattice thermal conductivity in the barrier regions which have a higher local Seebeck coefficient, results in a larger overall increase in the Seebeck coefficient.[21,55,65] We have not investigated this effect here, however, it might be the case that such an effect might not be utilized strongly for NIs as their filtering capabilities are weaker.

**<u>Diffusive to ballistic scattering conditions:</u>** The structures studied up to this point have used a mean-free-path (mfp) for scattering of mfp = 15 nm and channel length $L$ = 60 nm, which resulted in transport being diffusive within the channel. In reality, different materials can have different mfp's, and materials with very light effective masses could even be ballistic in the relatively short channel we simulate. Thus, to cover these cases, in Fig. 9 we investigate the main outcomes when channels with different transport regimes are considered: i) ballistic transport (black lines), ii) a channel of larger mfp of 30 nm (blue lines), and iii) a channel with smaller effective mass (green lines). In Fig. 9a, 9b and 9c we show the conductance, Seebeck coefficient and power factor, respectively for the 8×4 hexagonal array of nanoinclusions of diameter $d$ = 3 nm and $E_F$ = 0.05 eV. With red lines we show the corresponding results with mfp = 15 nm and m* = m$_0$ (same as the red lines in Fig. 2 and Fig. 3).

It can be seen that variations in both the mean-free-path and the effective mass have some effect on $G$ especially for low $V_B$, but their importance is reduced for larger barrier heights. The effect on $S$, on the other hand is negligible for low $V_B$ because the energy of the current flow does not change at first order with mfp or effective mass. Consequently the *PF* is benefitted by ~50% when doubling the mfp's or halving the channel material effective mass, which is expected. Importantly, qualitatively the trend for both cases is very similar to what was seen before. This consistency in the behavior can also be seen from the transmission shown in the inset of Fig. 9b. This shows that the transmission features vary only marginally between the three cases, with the lighter mass and larger mfp channels having a somewhat larger transmission. These results appear to show that the qualitative trends presented previously are robust to variations in mean-free-path and effective mass, suggesting that our conclusions could be applicable to a wide variety of material cases.

In the ballistic case (black lines), while $G$ and the $PF$ increase significantly compared to the diffusive case, it should be emphasized that even here all three parameters follow the trends previously outlined. The fact that $S$ is lower for low $V_B$ has to do with the shape of the transmission (black line in the inset of Fig. 9b), rising faster at lower energies, contributing a greater number of lower energy electrons to the current flow than seen in the diffusive channels.

**Approximations and omissions:** Finally, in this work, for computational simplicity, we have applied the NEGF formalism to short channel 2D nanoribbon structures of sizes $W = 30$ nm and $L = 60$nm embedded with hexagonally placed nanoinclusions and showed how it is a very powerful method that captures most of the important details for the assessment of the power factor. In reality, most of the experimentally realized structures are in 3D, which would have made our simulations computationally prohibitive. However, qualitatively, we believe our conclusions still apply to 3D structures as well. In fact, in Ref. 66 we considered the influence of random variations in the placement of barriers in superlattices, and found that it makes no difference in the power factor; thus we expect the main conclusions to qualitatively carry over from a regular set of structures to a more randomized colloidal placement with only an average separation as well. Furthermore, extrapolating from our findings, we expect that the influence of nanoinclusions on energy filtering in 3D would be even smaller since many more paths exist for the carriers to flow around the nanoinclusions.

In addition, a certain number of approximations have been made that we would like to elaborate on. First, the nanoinclusions were treated in a very simplified way, just by raising the potential barrier locally. Although this at first order can mimic a nanoinclusion, in reality material deformations exist in the vicinity of the inclusion, strain fields are built, and the effective mass and band details vary, which could have some quantitative influence on our results. Another omission is that in this work we have only considered electron-acoustic phonon scattering (in addition to the electron scattering on nanoinclusions).

Optical phonon scattering provides energy relaxation and it is important for optimizing energy filtering in superlattices where electrons absorb phonons to flow over potential barriers and emit phonons in order to relax into the wells.[55,66] In this case for nanocomposites, however, where most of the charge flows around the nanoinclusions, we omit optical phonons. The inclusion of optical phonons requires an additional computational complexity, which we relax in the interest of being able to simulate larger geometries that more elucidate the effect of nanoinclusions. Electron-ionized impurity scattering is an important mechanism, especially in degenerately doped materials, which can also result in a different energy dependence of the transmission function. In general, although ionized impurity scattering results in significantly lower power factors to begin with, the stronger energy variation in the transmission provides larger opportunities for relative power factor improvements,[58] thus, we expect that if that was included in our calculations the power factors would be qualitatively lower, but the increase nanoinclusions could provide would have been somewhat higher. In the case of energy filtering over a barrier in a superlattice, for example, we have previously shown that under ionized impurity scattering power factor improvements could reach up to 30%-40%,[21,58] whereas the relative improvement is half of that when only electron-phonon scattering is considered in the calculations.

Finally, we also need to mention that the perfect barrier shape we employ is just an approximation for ease in limiting the number of simulations to be performed and for focusing on the effects of geometry and density. In reality, in the vicinity of the heterojunction there will be potential variations that will affect the shape of the barrier, which as we show in Refs. 64 and 65, could be important in determining the PF. These potential variations are determined by the junction details, but also by the doping of the different regions and could only be captured accurately through self-consistent calculations, which we do not consider in this work. Figure 10 illustrates various cases of how the barrier shape will look once self-consistent electrostatics is considered (in this case through 1D simulation). In Fig. 10a we show the perfectly square barrier we use in the simulations as inserted in the Hamiltonian 'by hand'. In Fig. 10b we show how the barrier will look like when uniform doping ($N_D = 1.37 \times 10^{20}$/cm$^3$ places the $E_F$ at $E_F = 0.05$eV) is applied in all domains – in that case Schottky barriers are formed around the nanoinclusion.

Figure 10c shows how the barrier around the nanoinclusion looks like when only the matrix material is doped, whereas the nanoinclusion remains undoped. Finally, Fig. 10d shows a case with variable doping where the doping in the nanoinclusion is reduced to 30% of that in the matrix material. In the latter case the barrier profile looks very similar to the one simulated. In each case, the barrier is of course different, and will impact the results. The important point here, however, is that through electrostatic and charging effects a different 'effective barrier' is produced and this is what we consider.

## V. Conclusions

In conclusion, using the fully quantum mechanical Non-Equilibrium Green's Function method, we calculated the thermoelectric power factor of 2D nanoribbon channels with embedded nanoinclusions modelled as potential barriers. We explain why this method is most relevant, as it captures all geometry details, important quantum mechanical effects such as tunneling and subband quantization, as well as relevant transport regimes from diffusive to ballistic, and coherent to incoherent. These are all important features that affect transport through such structures and need to be captured for an accurate understanding of their thermoelectric properties as we showed in the results throughout the paper. Thus, this work avoids approximations in geometry and in essential transport features that could limit the proper design and optimization of nanostructured thermoelectrics. An important message of the paper is that we showed that, unfortunately, the addition of nanoinclusions does not utilize energy filtering effectively, and cannot provide higher power factors compared to an optimized structure without nanoinclusions (in the optimal pristine material case the Fermi level is placed around the conduction band edge). The introduction of nanoinclusions reduces the conductance, but does not provide the corresponding increase in the Seebeck coefficient. However, under degenerate conditions, where the Fermi level is placed into the conduction band, moderate increases in the power factor can be achieved of the order of 5-10% if the nanoinclusion barrier heights are between the Fermi level $E_F$ and the conduction band $E_C$. Importantly, however, we showed that in that case, the mild power factor improvements are independent of the nanoinclusion density, as long as the nanoinclusions are large enough to prevent quantum tunneling. This indicates that larger densities of relatively thick nanoinclusions can be utilized to effectively reduce the lattice thermal conductivity without degradation in the power factor. For larger barrier heights, a non-intuitive simultaneous drop in the conductance and Seebeck coefficient is observed, which degrades the power factor significantly. Our results reveal that the filtering behavior of materials with nanoinclusions are different compared to the filtering behavior of cross-plane superlattices. Our conclusions would be useful in the design of advanced nanostructured thermoelectric materials.

Acknowledgements: This work has received funding from the European Research Council (ERC) under the European Union's Horizon 2020 Research and Innovation Programme (Grant Agreement No. 678763). M.T. has been supported by the Austrian Research Promotion Agency (FFG) Project No. 850743 QTSMoS.

Figure 1:

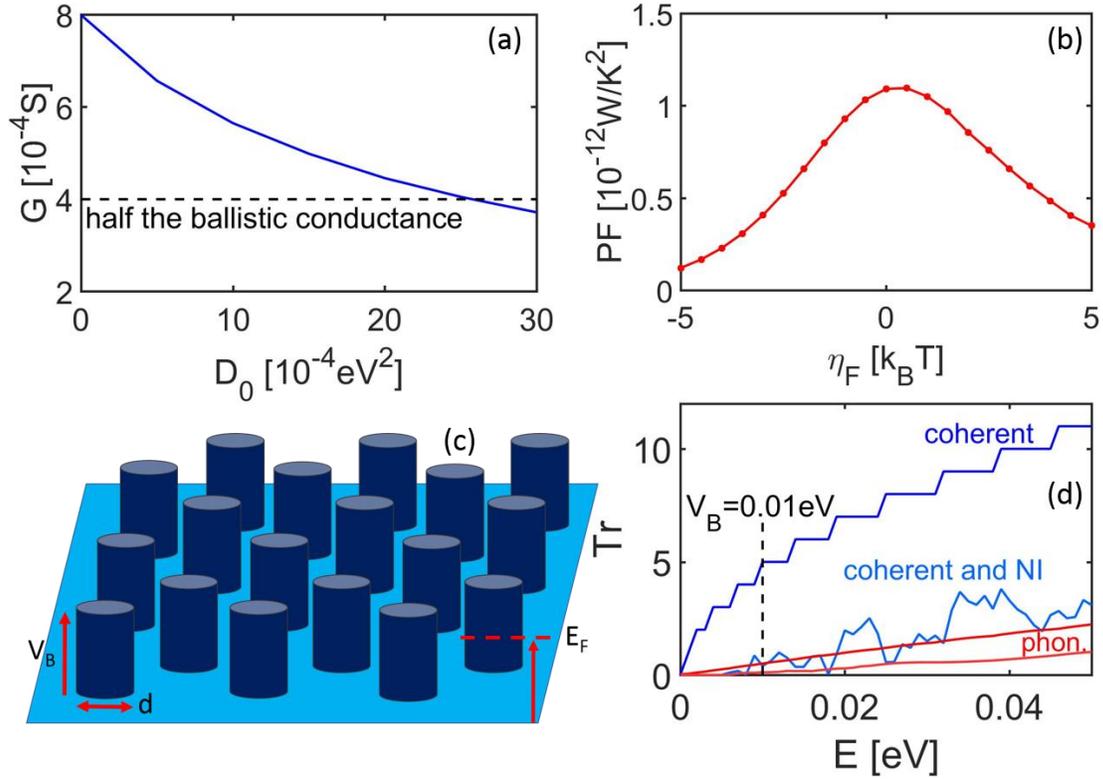

## Figure 1 caption:

(a) Calibration of the simulations' scattering parameters. The scattering strength is increased in an $L = 15$ nm channel until the conductance falls to half of the ballistic value (dashed-black line), thereby setting the mean-free-path of the electrons to 15 nm. (b) The power factor (defined as $GS^2$) of a pristine (without nanoinclusions) channel as the Fermi level is scanned across the bands. (c) A schematic of a typical geometry we consider. $V_B$ is the barrier height, $d$ the nanoinclusion diameter, and $E_F$ the Fermi level. (d) A comparison of the transmissions for an empty channel under ballistic coherent conditions (blue line), a channel with nanoinclusions under coherent transport (light-blue line), an empty channel under phonon scattering transport conditions (red line), and a channel with nanoinclusions under phonon scattering transport conditions (light-red line).

Figure 2:

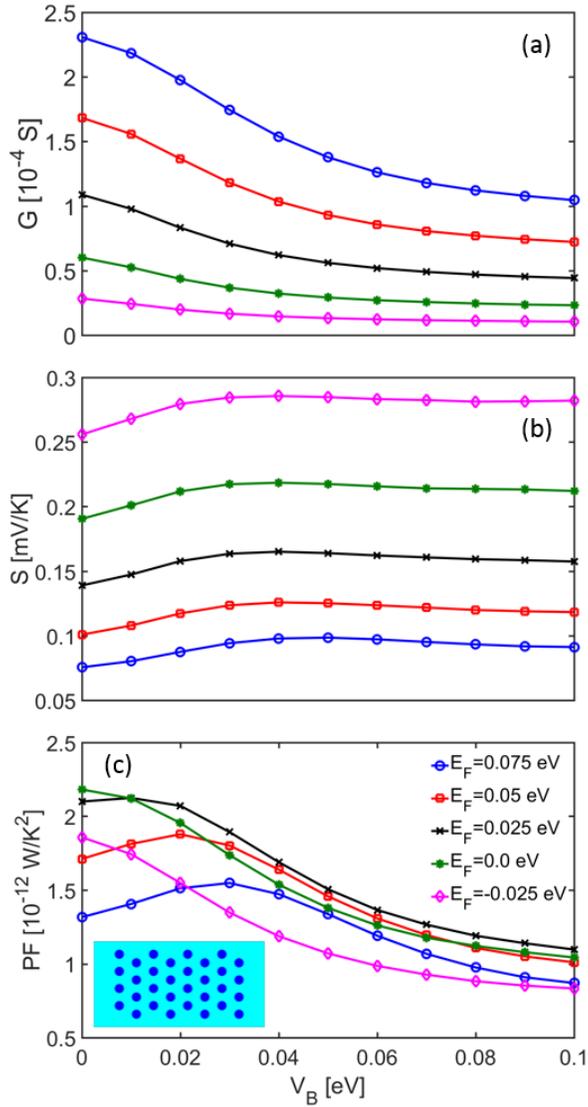

Figure 2 caption:

The thermoelectric coefficients of an $L$ = 60 nm channel with an 8×4 hexagonal arrangement of nanoinclusions (inset of (c)) and acoustic phonon scattering transport conditions versus nanoinclusion barrier height, $V_B$. (a) The conductance. (b) The Seebeck coefficient. (c) The power factor defined as $GS^2$. Five different Fermi levels are considered: $E_F$ = -0.025 eV (purple-diamond lines), $E_F$ = 0 eV (green-star lines), $E_F$ = 0.025 eV (black-cross lines), $E_F$ = 0.05 eV (red-square lines), and $E_F$ = 0.075 eV (blue-circle lines).

Figure 3:

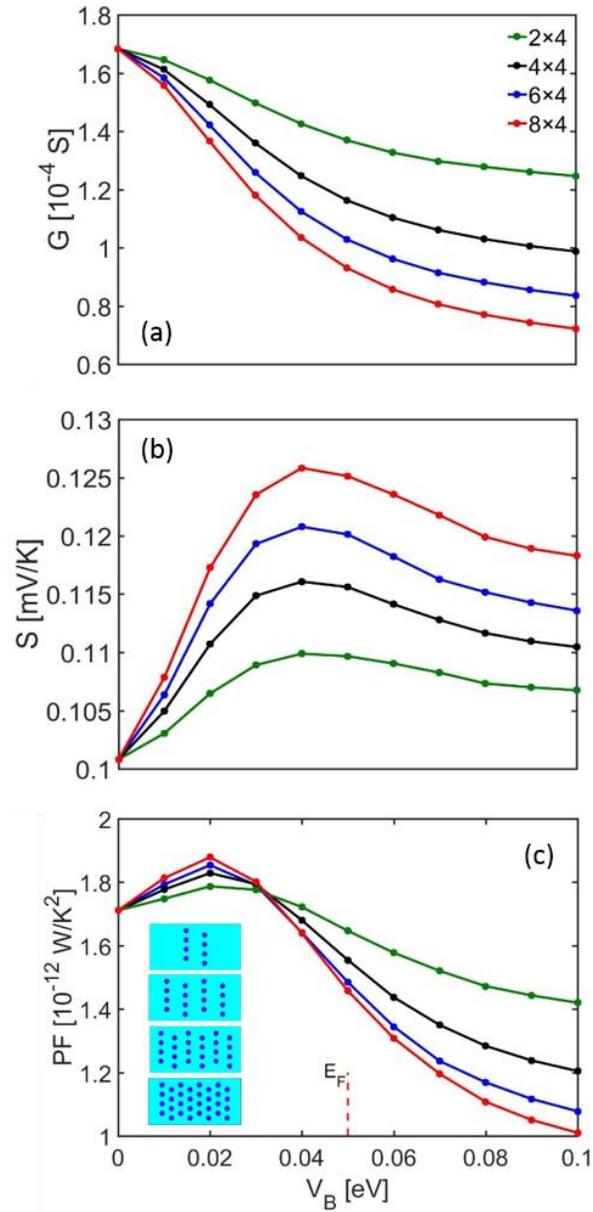

Figure 3 caption:

The thermoelectric coefficients of an $L = 60$ nm channel with $E_F = 0.05$ eV (dashed-red line) and acoustic phonon scattering transport conditions versus nanoinclusion barrier height, $V_B$. (a) The conductance. (b) The Seebeck coefficient. (c) The power factor defined as $GS^2$. Hexagonal arrays of four different nanoinclusion densities are considered as shown in the inset of (c): 2×4 array (green lines), 4×4 array (black lines), 6×4 array (blue lines), and 8×4 array (red lines).

Figure 4:

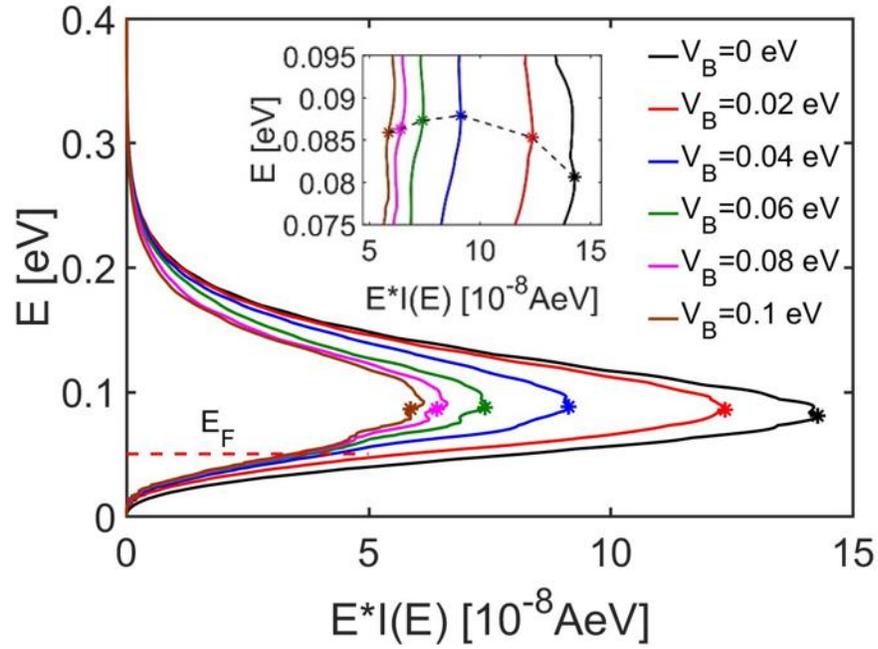

Figure 4 caption:

The distribution of the energy of the current flow for an $L = 60$ nm channel with an 8×4 array of nanoinclusions and $E_F = 0.05$ eV. The stars denote the average energy of the current flow. A zoomed version of these is shown in the inset. Six different nanoinclusion barrier heights are shown: $V_B = 0$ eV (black), $V_B = 0.02$ eV (red), $V_B = 0.04$ eV (blue), $V_B = 0.06$ eV (green), $V_B = 0.08$ eV (purple), and $V_B = 0.1$ eV (brown). The dotted line in the inset indicates from right to left the trend of increase in $V_B$.

Figure 5:

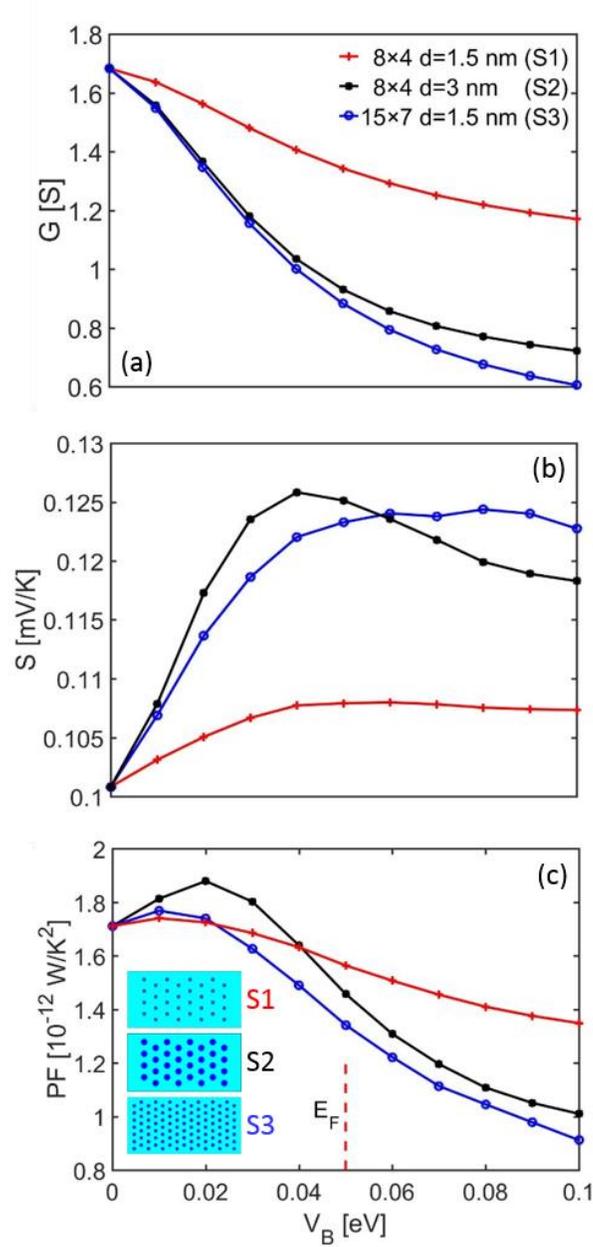

Figure 5 caption:

The thermoelectric coefficients of $L$ = 60 nm channels (insets of (c)) with 8×4 array of nanoinclusions versus nanoinclusion barrier height, $V_B$, for two different nanoinclusion diameters: $d$ = 1.5 nm (red lines) and $d$ = 3 nm (black lines), and a 15×7 array with $d$ = 1.5 nm (blue lines) whose density is equivalent to the 8×4 array with $d$ = 3 nm. (a) The conductance. (b) The Seebeck coefficient. (c) The power factor defined as $GS^2$.

Figure 6:

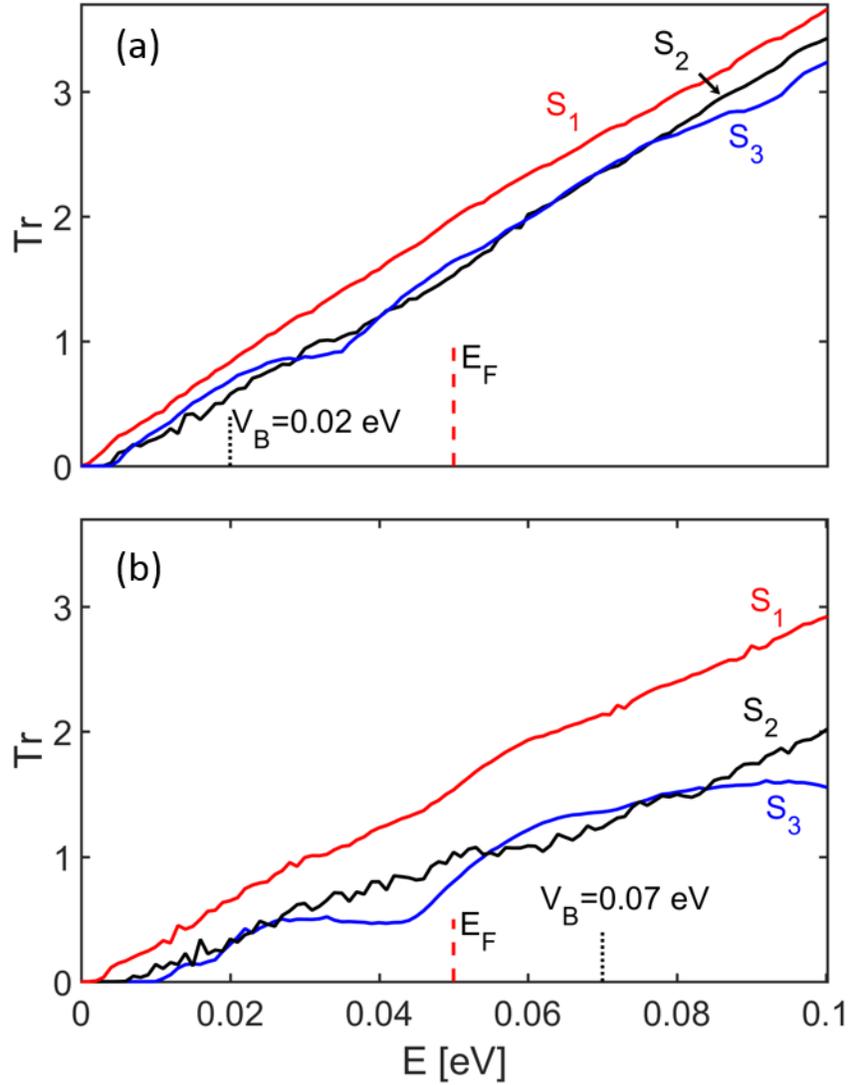

**Figure 6 caption:**

The transmission versus energy for the channels shown in the inset of Fig. 5c: 8×4 arrays of $d$ = 1.5 nm (red lines) and $d$=3 nm (black lines), and a 15×7 array with $d$ = 1.5 nm whose density is equivalent to the 8×4 array with $d$=3 nm (blue lines). (a) The case for nanoinclusions with $V_B$ = 0.02 eV barrier height, i.e. before the $E_F$, where the power factor reaches the maximum point in Fig. 5c. (b) The case for nanoinclusions with $V_B$ = 0.07 eV barrier height, i.e. after the $E_F$, where the power factor starts to saturate in Fig. 5c.

## Figure 7:

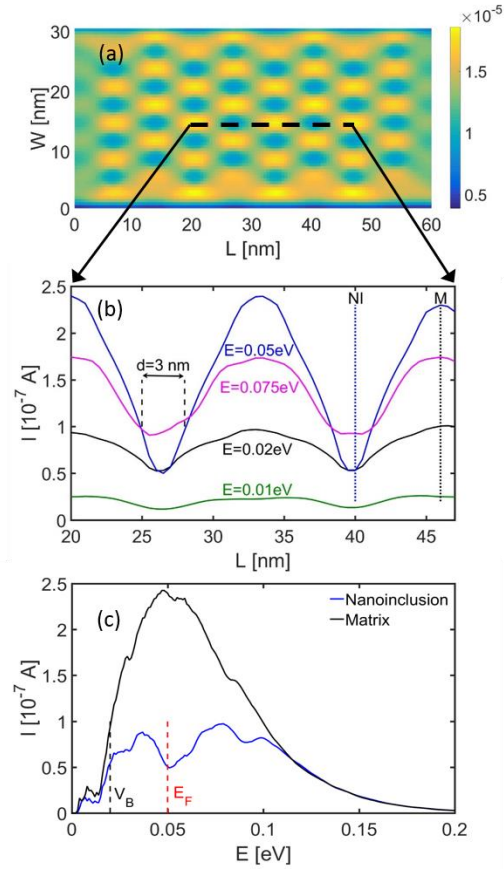

## Figure 7 caption:

(a) Colour map of the current flow directed along the length of the channel (*L*-directed) through an 8×4 hexagonal array of nanoinclusions ($d = 3$ nm, $V_B = 0.02$ eV). Nanoinclusions can be seen as the blue areas and the matrix material as the yellow and green areas. (b) The channel length directed current along the dashed-black line shown in (a) at four different energies, below the barrier at $E = 0.01$ eV (green line), at the barrier $E = 0.02$ eV (black line), at the Fermi level $E = 0.05$ eV (blue line), and above the barrier and Fermi level at $E = 0.075$ eV (purple line). The location of the first nanoinclusion (NI), which extends for 3 nm, is denoted. (c) The current flow at two points in the structure: at the centre of a nanoinclusion (blue line, position shown by dotted-blue line in (b) at $L \sim 40$ nm) and in the pristine matrix (black line, position shown by dotted-black line in (b) at $L \sim 46$ nm). The barrier height is shown by the dashed-black line and Fermi level by the dashed-red line.

Figure 8:

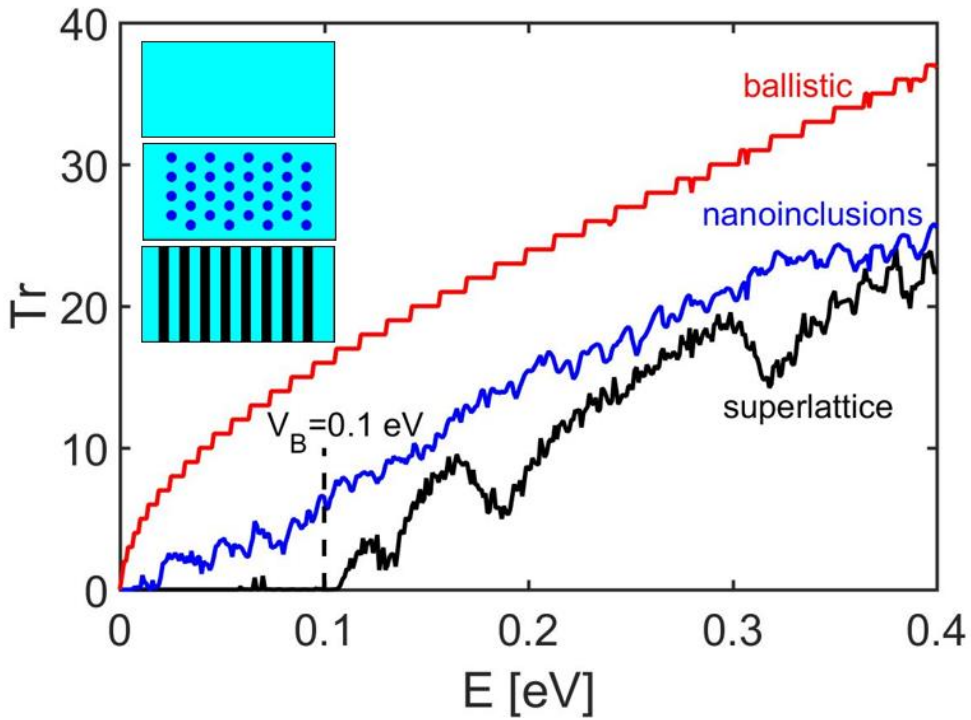

Figure 8 caption:

The transmission versus energy for an $L = 60$ nm ballistic coherent channel (no phonon scattering) for the following cases as shown in the insets: i) pristine material without nanoinclusions (red line), ii) material with an 8×4 hexagonal array of nanoinclusions (blue line) with $V_B = 0.1$ eV, and iii) a superlattice material (black line) with $V_B = 0.1$ eV. The barrier height $V_B$ is marked by a dashed-black line. It can be seen that the superlattice is effective at cutting out the contribution of low energy electrons (achieving an increase in the Seebeck coefficient) whereas the nanoinclusions act to reduce the transmission uniformly in the entire energy region.

## Figure 9:

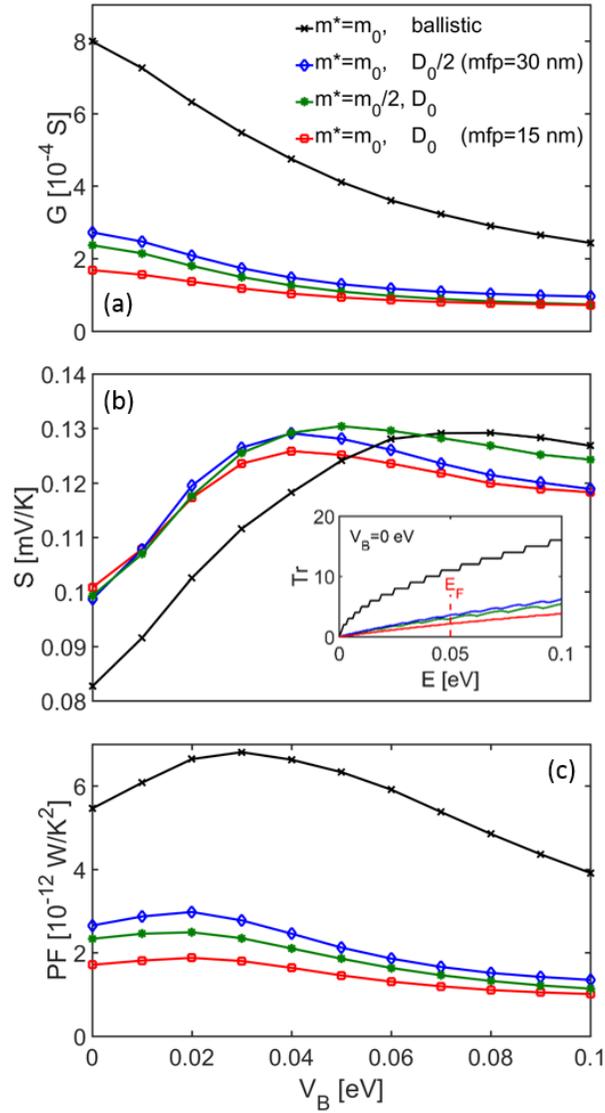

## Figure 9 caption:

The thermoelectric coefficients of $L = 60$ nm channels with an 8×4 array of nanoinclusions versus nanoinclusion barrier height, $V_B$, for four different simulation conditions: Ballistic transport (black lines), mean-free-path mfp = 30 nm and m* = m$_0$ (blue lines), mean-free-path mfp = 15 nm and m* = 0.5m$_0$ (green lines), and mean-free-path mfp = 15 nm and m* = m$_0$ (red lines − same as in Fig. 2 and Fig. 3). (a) The conductance. (b) The Seebeck coefficient. (c) The power factor defined as $GS^2$. Inset of (b): The transmission probability versus energy in the four cases for $V_B$=0 eV.

Figure 10:

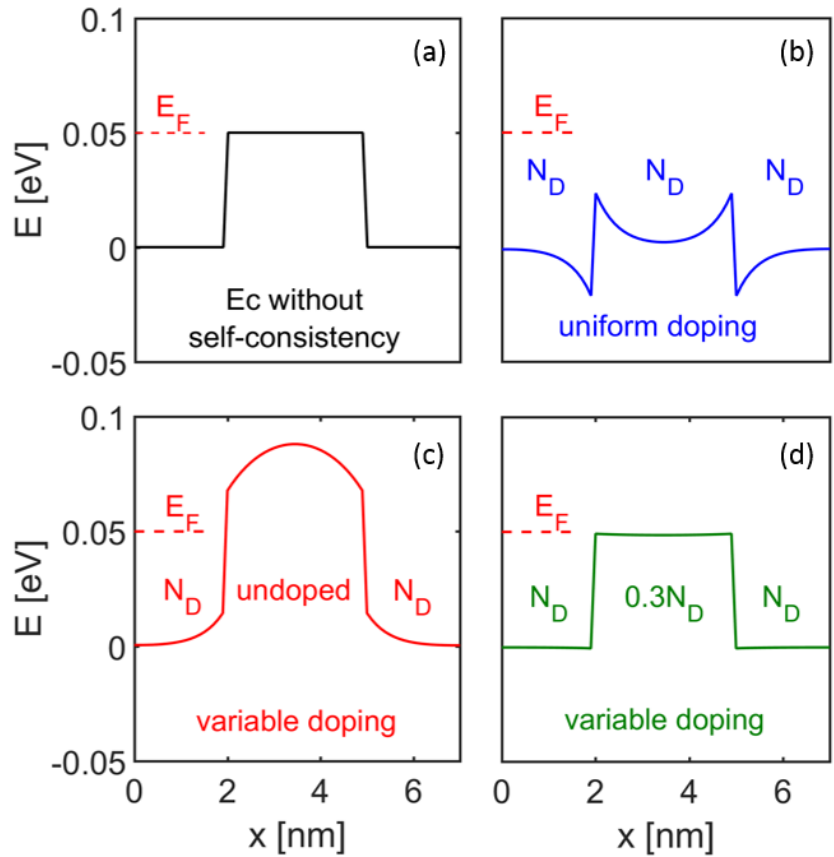

Figure 10 caption:

The shape of the barrier around the nanoinclusion for different cases using 1D self-consistent calculations. (a) The perfectly square barrier as used in the simulations. (b) The barrier shape when uniform doping is applied in all domains – Schottky barriers are formed around the nanoinclusion. (c) The barrier shape around the nanoinclusion when only the matrix material is doped, whereas the nanoinclusion remains undoped. (d) A case with variable doping where the doping in the nanoinclusion is reduced to 30% of that in the matrix material. In the latter case the barrier profile looks very similar to the one simulated.